\UseRawInputEncoding
\documentclass[nofootinbib, twocolumn,superscriptaddress,prl]{revtex4-1}

\usepackage{epsfig}
\usepackage{graphicx}
\usepackage{amsmath}
\usepackage{amssymb}
\usepackage{epstopdf}
\usepackage{xcolor}
\usepackage{subfigure}
\usepackage{bm}
\usepackage{dsfont}
\usepackage{hyperref}
\usepackage{braket}
\usepackage{todonotes}
\usepackage{physics}
\usepackage{tikz}
\usepackage{pgfplots}
\newcommand{\bea}{\begin{eqnarray}}
\newcommand{\eea}{\end{eqnarray}}

\begin{document}

 \title{Detecting Acceleration-Enhanced Vacuum Fluctuations with Atoms Inside a Cavity}

\author{Kinjalk Lochan}
 % \email{kinjalk@iisermohali.ac.in}
 \affiliation{Department of Physical Sciences, Indian Institute of Science Education and Research (IISER) Mohali, Sector 81 SAS Nagar, Manauli PO 140306, Punjab India.}

\author{Hendrik Ulbricht}
% \email{h.ulbricht@soton.ac.uk}
\affiliation{School of Physics and Astronomy, University of Southampton, Southampton SO17 1BJ, United Kingdom}

 \author{Andrea Vinante}
 % \email{a.vinante@soton.ac.uk}
\affiliation{Department of Physics and Astronomy, University of Southampton, Southampton SO17 1BJ, United Kingdom}
\affiliation{Istituto di Fotonica e Nanotecnologie - CNR and Fondazione Bruno Kessler, I-38123 Povo, Trento, Italy}

 \author{Sandeep K.~Goyal}
 \email{skgoyal@iisermohali.ac.in}
 \affiliation{Department of Physical Sciences, Indian Institute of Science Education and Research (IISER) Mohali, Sector 81 SAS Nagar, Manauli PO 140306, Punjab India.}

\begin{abstract}
  Some of the most prominent theoretical predictions of modern
times, e.g., the Unruh effect, Hawking radiation, and gravity-assisted
particle creation, are supported by the fact that various quantum constructs like particle
 content and vacuum fluctuations  of a quantum field are observer-dependent. Despite being fundamental in nature,
 these predictions have not yet been experimentally verified because %\remove{ remain to be experimentally  verified till date},  because
one needs extremely strong gravity (or acceleration) to bring  them within the existing experimental resolution. In this Letter, we  demonstrate 
that a post-newtonian rotating atom inside a far-detuned
cavity experiences strongly modified quantum fluctuations in the 
inertial vacuum. As a result, the emission rate of an excited atom gets
enhanced significantly  along  with a shift in the emission spectrum due to the change  in the quantum correlation under rotation. We propose an optomechanical setup that is capable of realizing such  acceleration-induced particle creation with current technology.
This   provides a novel and potentially feasible experimental proposal for the direct detection of noninertial quantum field theoretic effects.

\end{abstract}

\maketitle
% \section{Introduction}
Random fluctuations  in vacuum are a well-known phenomenon in quantum field theory that manifests in various places. e.g., Casimir effect, Hawking radiation and the  Schwinger effect etc.~\cite{Casimir1948,Hawking1974,Schwinger1951}.   Typically of tiny strength, these random fluctuations can be amplified in extreme physical scenarios such as  strong gravitational field~\cite{Fulling1973,Davies1976} such that they start appearing as real particles. Because of the equivalence principle,  a similar amplification in these fluctuations can also be observed in highly accelerating frames like  the Unruh effect states in which  a uniformly accelerating observer perceives the inertial vacuum as thermally populated~\cite{Unruh1976}. Therefore, the particle content of a quantum field is truly observer- and frame-dependent ~\cite{Fulling1973,Hawking1974,Davies1976}.

An operational approach to detecting the presence of particles in a quantum state is through a Unruh-Dewitt detector (UDD)~\cite{Unruh1976}, a proposed quantum device in which a  quantum mechanical system couples with a permeating  quantum field and registers transitions if it sees the field in a nonvacuum state. Such a detector, if put on a noninertial trajectory, indeed detects particles in terms of clicks (number of transitions in the quantum system) even when the state of the field is vacuous according to inertial observers.  The response of UDD truly couples to the correlation function rather than true field content \footnote{In standard literature, true field content in any frame relates to local observables e.g., number of particles, stress energy expectation, etc. All of these are of course derivable from the correlation function (as well as  its derivatives). But still there exist cases in which the field content is zero, yet the detector clicks as it only couples to correlator and not to its derivatives  ~\cite{Davies1996}.}. Yet if a detector were to pick up the minute changes in the correlators due to acceleration, it would be very helpful in characterizing the cases where the field content is truly changed under noninertial motion such as the Unruh effect ~\cite{Unruh1976}. 
Moreover, one can envision that there could be other features of a UDD that could be equally (or even more) sensitive to such changes and respond strongly when put in  noninertial motion. For example, if the detector is in the excited state, the rate of its  ``deexcitation'' is very much sensitive to changes in the quantum correlators in the 
environment in which it is coupled.

Despite a robust theoretical genesis, such noninertial transitions remain experimentally unverified, primarily because any appreciable noninertial effect would demand ridiculously high acceleration. For example,  uniform acceleration makes the UDD response perfectly  thermal, but any observable thermal signature will require a uniform acceleration as large as  $10^{26} ms^{-2}$~\cite{Crispino2008}. There have been attempts to enhance such effects % to make them worthy of observation
using techniques such as ultraintense lasers~\cite{Chen1999}, Penning traps~\cite{Rogers1988} and optical cavities~\cite{Scully2003} to obtain momentarily high accelerations  in order to witness any alteration in the response pattern of a quantum system. The approaches suggested for capturing the finite temperature effects of an accelerating system range from monitoring thermal quivering~\cite{Raval1996},  decay of accelerated protons ~\cite{Vanzella2001}, or even radiation emissions~\cite{Retzker2008, Schutzhold2006} from such systems. Also, other quantum features such as the judicious selection of Fock states~\cite{Aspachs2010} and using geometric phases~\cite{Martin-Martinez2011} have been suggested as ways to enhance the effect of noninertial motion. Unfortunately, all these efforts are still far away from being realized. Any observation of such noninertial distortions of the field correlators is not only important in itself but also reinforces our understanding of phenomena such as gravitational particle creation.

Constant acceleration is not the only way to observe the changes in field observables due to noninertial motion~\cite{Kothawala2010}. Any UDD launched on a general nongeodetic trajectory, e.g. a circularly rotating one, may also register {\it particles} ~\cite{Davies1996}. Interestingly, the rotating UDD does not click in the inertial vacuum for small rotational frequencies but registers the presence of  particles for rapid rotations the UDD response is caused by the change in the correlation function rather than the presence of actual particles \cite{Crispino2008}. (The inertial and the rotating set of modes are related through  vanishing Bogoliubov coefficient $\beta_{\bf kk'}$, thus agreeing on the vacuum state and field content ~\cite{Crispino2008}).

In this Letter, we propose a setup to facilitate the observation of such a  noninertiality driven UDD response via measuring the transition rates of a rotating atom inside an electromagnetic cavity. Once the atom achieves a sufficiently high rotation, the  particles ``perceived''   by the UDD \cite{Davies1996} also affect its other responses e.g. the change in the emission  rate if the UDD is in excited state. This is somewhat similar to stimulated emissions or  cases involving a true  change in field content in noninertial motion e.g.,  the Unruh effect. Inside a cavity, the modified boundary conditions naturally enhance the inertial spontaneous emission, and its resonace frequency provides an additional means to capture the emissions assisted by noninertial motion occurring at a frequency different from the natural transition line of the atom. We show that in the leading order of the 
post-Newtonian relativistic parameter $v^2/c^2 \sim R^2\omega^2/c^2$ (as it also signifies the change in spacetime metric), where $R$ is the radius of rotation and $\omega$ its angular frequency, the transition rate between suitably selected energy levels of an atom rotating inside a far-detuned  electromagnetic cavity can be significantly increased by controlling $\omega$ and choosing the parameters of the cavity  appropriately.  This  makes 
our setup one of the most sensitive proposals for  measuring noninertial quantum field theoretic effects. Admittedly this setup does not really measure the standard Unruh effect, but any detection of radiation caused by noninertial motion, even if it is nonthermal, substantiates the idea that the inertial vacuum is {\it perceived} differently by noninertial observers.

Let us consider an atom with two energy levels separated by energy $\Delta E = \hbar\Omega$. Due to its electronic configuration, the atom possesses a well-defined electric dipole moment, say $\hat d^\mu$, that will interact with the electric field of the surroundings. % , which are prone to interact with the dipole moment of the atom.
Hence, the interaction between the atom and the surroundings is $H_I = -\hat d^\mu E_\mu$, where $E_\mu$ represents the electric field operator~\cite{Anandan2000}. This interaction  can cause a transition between the excited state and the ground state of the atom (spontaneous emission). The  transition rate can be calculated using the two point function of the electric field 
in any given quantum state.
For an atom at rest interacting with the surrounding electric field  in a vacuum state (as per the laboratory inertial frame), the transition rate  is given by $\Gamma_0 = \Omega^3 {d}^2/3\pi\epsilon_0\hbar c^3$, where ${d}^2 = |\langle \psi_f |\hat d | \psi_i\rangle|^2$  (see appendix).

 {\em Electric dipole coupling in the accelerated frame.}
 If an atom is set in an eternally accelerating motion, the rate of transition and the spontaneous emission both will  become thermal, as expected~\cite{Unruh1976}. In the Unruh effect, an atom moving with uniform acceleration $a$ experiences a temperature $T = \hbar a/2\pi  k_B c$ which results in the modified transition rate $\Gamma_T = \Gamma_0 /(1-\exp{-2 \pi c \Omega /a})$.
 Any appreciable effect of the acceleration-induced temperature will require $ a \sim 2\pi c \Omega$; however, achieving such higher acceleration scales, for typical atomic transitions is beyond the reach of current technology.

In order to observe such noninertial effects, one may consider reducing $\Omega$ to very low values, e.g., radio frequencies ($10^7$ Hz). 
Unfortunately, at such low frequencies,  the natural spontaneous emission rates ($\Gamma_0$) of typical atoms are so vanishingly small that any hope of seeing the noninertial modifications even at  high accelerations, becomes very feeble. If by any means we increase $\Gamma_0$ while keeping $\Omega$ low then the noninertial effects can also possibly  be observed efficiently. An electromagnetic cavity is a well-known tool  to achieve this goal \cite{purcell}.  The electromagnetic cavity modifies the boundary conditions for the surrounding field, which leads to a change in density of the states of the electric field modes. This in turn  alters the transition rate of the atom.  In a cavity of volume $V$ and quality factor $Q$ the inertial spontaneous emission rate at resonance is given by
\bea
\Gamma_{cav,in} \sim \frac{d^2}{\epsilon_0 \hbar} \frac{1}{V} Q,\label{Eq:Gamma_c}
\eea
which is independent of the transition frequency $\Omega$.
Next we show that a rapidly rotating quantum system inside an electromagnetic cavity, displays an enhanced transition rate.

{\em Acceleration-induced emission inside cavity.} 
 The proposed setup consists of an atom rotating in a circular trajectory, %of radius $R$ and frequency $\omega$ inside an electromagnetic cavity. The 
with the direction of the atomic transition dipole $\hat{d}^{\mu}$ taken as orthogonal to the plane of rotation.
The relevant expression for the spontaneous emission rate,  if we do the expansion for the time averaged transition rate (over many cycles) in the leading order of $\zeta(\omega) \equiv R^2 \omega^2/c^2$, reads as follows (see appendix):
\begin{widetext}
\bea
\Gamma_{cav} \sim  \frac{d^2}{\epsilon_0 \hbar} \frac{1}{V}\int_0^{\infty} dk  \rho(k)\omega_k \left[  \delta(\bar{\Omega} -\omega_k) +\frac{3}{2}\frac{ R^2}{c^2} \left(\frac{\omega^2}{3}+ \frac{\omega_k^2}{30}  \right)\{\delta(\bar{\Omega} + \omega -\omega_k )+ \delta(\bar{\Omega}  - \omega - \omega_k )\} \right.\nonumber\\
\left.   - \frac{1}{10}  \frac{ R^2 \omega_k^2}{c^2} \delta(\bar{\Omega}  -\omega_k ) +{\cal O}\left(\sum_{n>1} \frac{R^{2n} \omega_k^{2n}}{c^{2n}}\delta(\bar{\Omega}  \pm n\omega - \omega_k )\right)\right],
\eea
\end{widetext}
with $\bar{\Omega}  = \Omega  \left( 1-\omega^2 R^2/c^2\right)^{1/2} \approx \Omega$.
where $\rho(k)$ is the density of field states of the electromagnetic modes inside the cavity. In the case of $\omega\gg \bar{\Omega}$,  the above expression can be approximated in the leading order of $\zeta$ to
\bea
\Gamma_{cav} \sim \frac{d^2}{\epsilon_0 \hbar V} \left[  \bar{\Omega} \left( 1-\frac{\zeta(\bar{\Omega})}{10} \right) \rho( \bar{\Omega})+ \frac{33\zeta(\omega)}{60}\tilde{\omega} \rho(\tilde{\omega} )\right] \label{CavityResponse}
\eea
with $\tilde{\omega}=(\omega+\bar{\Omega})$.  In cavities, boundary effects, that lead to sharp transients as well as modified density of states affect the inertial responses, too, in order to increase the rates substantially \cite{Obadia:2007cg, Crispino2008}. Boundary effects are not expected to play a significant role in the rotating case, if one is sufficiently distant  from the boundaries, by choosing  rotor sufficiently smaller than the cavity size. Still, to clearly isolate the effects caused  solely by  rotation, we can remove the inertial response inside the cavity from the total emission rate as follows:
\bea
\Delta \Gamma_{cav,ni} =  \Gamma_{cav}(\omega) -\underbrace{\Gamma_{cav}(\omega \rightarrow 0)}_{\Gamma_{cav,in}}, 
\eea 
resulting in
\bea
\Delta \Gamma_{cav,ni} =   \frac{d^2 \zeta(\omega)  \Omega}{2 \epsilon_0 \hbar  V}  \left[ -\rho(\Omega)- \Omega\rho'(\Omega)+ \frac{33}{30}\frac{\tilde{\omega}}{\Omega} \rho(\tilde{\omega} )\right], \label{CavityResponse2} 
\eea
up to ${\cal O}(\zeta)$.

\begin{figure}
\includegraphics*[width=1.0\columnwidth]{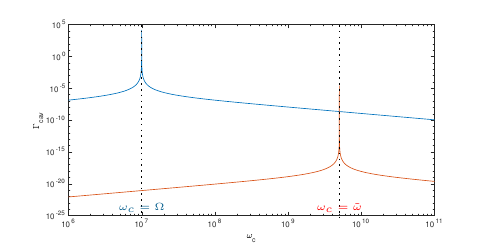}
\caption{The transition rate inside the cavity including the inertial (blue curve) and noninertial (red curve) contribution in log scale.}
\label{fig:RotationContribution}
\end{figure}

The density of states inside the cavity is typically taken as a Lorentzian profile of width $\omega_c/Q$
\bea
\rho(\omega_k) \sim \frac{(\omega_c/Q)}{(\omega_c/Q)^2 + (\omega_k -\omega_c)^2}
\eea
with the quality factor $Q \gg 1,$ and $\omega_c$ being the normal mode frequency of the cavity. 
We see that if we put $\omega_c \sim \Omega$, i.~e., when the cavity is in resonance with the atomic transition, the first term in Eq.~\eqref{CavityResponse} dominates and we obtain an expression close to the standard spontaneous rate inside the cavity as given in Eq.~\eqref{Eq:Gamma_c}.
 However,  if $\omega_c \sim \tilde{\omega}$ then the second term of Eq.~\eqref{CavityResponse} dominates the standard term by many orders. Similarly in Eq.~\ref{CavityResponse2}, the third term becomes dominant over other two terms in this case and the total rate of transition can be approximated to
\bea
 \Gamma_{cav} & \longrightarrow & \Delta \Gamma_{\text{cav,ni}}  \sim  \frac{33\zeta(\omega)}{60} \frac{d^2}{\epsilon_0 \hbar} \frac{1}{V} Q , \label{NIResponse}\\
q (\omega_c)&\equiv& \frac{\Delta \Gamma_{\text{cav,ni}}}{\Gamma_{\text{cav, in}}} = \frac{33\zeta(\omega)\tilde{\omega}\rho(\tilde{\omega})}{60 \Omega \rho(\Omega)}\rightarrow \frac{\omega_c}{\Omega}Q^2 \zeta(\omega_c),
\eea
as $\omega_c$ approaches $\tilde{\omega}$. Thus,  when the cavity is tuned near $\tilde{\omega}$ and $Q^2 \zeta > \Omega/\omega_c$, the contribution of the noninertial motion becomes dominant, see Fig.~ \ref{fig:RotationContribution}.

If the parameters inside the cavity could be arranged to the values
$\omega_c \sim GHz$, $\zeta \sim 10^{-13}, d\sim 10^{-29}\text{Cm}, V\sim 10^{-14}\text{m}^3$ with a quality factor $Q\sim 10^7$ we would have a transition rate of the order of $\Gamma_{cav} \sim 10^{-7} s^{-1}$. For a system of, say, $10^6$ atoms, we can expect hundreds of events per hour. As a comparison, in this far-detuned region, the contribution from the inertial term is $\Gamma_{cav} \sim 10^{-10} s^{-1}$ with $\Omega \sim 10 $ MHz and hence $q \sim10^4$. Furthermore, this transition would be mediated by photons with frequency $\omega_c$, which is very different from the natural transition frequency $\Omega$. Observation of the photons with frequency $\omega_c$ itself would be a testimony to the effect of noninertial motion on the quantum field. In the following, we present the experimental feasibility of our proposal. 

\begin{figure}
\includegraphics[width=1.0\columnwidth]{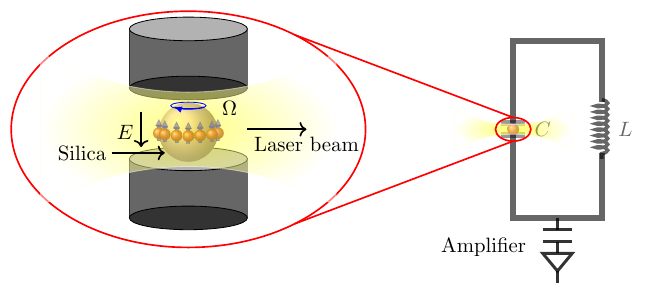}
\caption{ A silica nanoparticle is levitated and spun at a frequency $\omega \approx 5$ GHz by a circularly polarized focused laser beam.
  The atoms placed on the silica particle feature a characteristic electric dipole allowed transition at frequency $\Omega \approx 10-100$ MHz. The nanoparticle is placed between the plates of a capacitor, which is part of a superconducting microwave LC resonator tuned at $\omega_c$. The noninertial motion of the atoms riding the silica particle will cause additional atomic transitions with mean frequency  $\omega+\Omega \approx \omega_c$. Emitted photons will be detected by a quantum limited microwave amplifier.}
\label{fig:scheme}
\end{figure}

{\em Proposed experimental setup.} 
  A schematic diagram for the proposed experimental setup is shown in Fig.~\eqref{fig:scheme}. 
In our setup, a silica particle (SiO$_2$) with radius $R\approx 50$~nm  is trapped by an optical gradient force in the focal spot of a focused optical beam that falls  inside the sensitive volume of a lumped element high $Q$ microwave cavity with frequency $\omega_c \approx 5$ GHz. The silica particle is then rotated with a frequency $\omega \approx \omega_c$ by means of circularly polarized light. Suitable atoms with the dipole allowed transition frequency $\Omega\approx 10$MHz are placed on top of the silica particle. 
The noninertial transitions, upconverted to $\omega+\Omega\approx \omega_c$, are detected by a quantum limited microwave amplifier. The whole experimental setup can be divided into four main components: (a) a levitated optomechanical oscillator on which we place our atoms, (b) the microwave cavity used to suppress undesired transition lines of the atom, (c) the appropriate atom with the desired range of transition frequencies, and (d) photon detector. Below, we present details of each of the experimental components.

{\it Levitated optomechanical oscillator:}  
A silica nanoparticle with radius $R\approx$50~nm will be trapped by the optical gradient force in the focal spot of a 1550~nm laser beam which also transfers its spin angular momentum to the trapped particle and causes it to rotate \cite{Kuhn2017}.  Introduction of this driving laser will of course produce additional resonances over and above the noninertial vacuum response, but these can be minimized by suitably selecting the cavity frequency away from that of the driving laser as well as using a narrow peak distribution of the driving field around  1550 nm with corresponding $\omega_{laser} \gg \omega + \Omega$ (see appendix).
Recent experiments have shown the ability to control the translational motion of trapped particles--for instance, by parametric feedback cooling~\cite{jain2016direct}, state preparation~\cite{frimmer2016cooling, setter2018real}, squeezing~\cite{rashid2016experimental} and an unprecedented potential for ultraprecise force sensing~\cite{geraci2010short, hebestreit2018sensing}.
 Beside the translational motion, different rotational motions have been experimentally demonstrated in levitated optomechanical systems such as libration~\cite{hoang2016torsional}, free rotation~\cite{arita2013laser, rahman2017laser}  and precession~\cite{rashid2018precession} of nonspherical nanoparticles.  The free rotation can be stabilized to utmost precision, $Q_{\text R}=10^{11}$~\cite{kuhn2017optically}, and can reach very large frequencies in the GHz range~\cite{reimann2018ghz, ahn2018optically}, which are only limited by the centrifugal damage threshold of the rotating particle. Here, we propose to use this high rotational frequency at the maximum experimentally demonstrated value $\omega$=5~GHz in order to observe the  emissions by the noninertial motion~\cite{ahn2018optically}.
  
{\it Microwave cavity:}  
A lumped LC resonator will be used as a microwave cavity~\cite{bertet}.  In addition, superconducting wiring is required for the cavity in order to achieve the highest possible electrical $Q$ factor. Typical resonators made on thin films at GHz frequency feature $Q$ factors up to $10^6-10^7$ \cite{Leduc2010}. The particle will be placed inside the capacitor gap, as shown in Fig.~\eqref{fig:scheme}. The capacitor could be made with the ends of two wires in front of each other, with the wires connected to a properly designed inductor. With a quasicubic geometry the microwave cavity could be designed with a resonator mode volume as low as $V \approx 10^{-14}$ m$^3$. The gap between the capacitor plates (of the order of $20\mu$m) will be large enough to accommodate the confining laser beam waist.
The LC resonator will be cooled to cryogenic temperatures to reduce the thermal spontaneous emissions. For temperatures around mK, it is expected to be in the quantum regime. The correction in emission rate  due to blackbody temperature at $T\sim m K$ for $\bar{\Omega} \sim 10 MHz$ will be  $\Gamma \rightarrow  \Gamma_0/(1-e^{-1/2}) \sim 2 \Gamma_0$, which is at least 2 orders smaller than the noninertial effect discussed in this Letter. Maintaining such a low  temperature for the levitated rotating nanosphere appears challenging, but estimations based on thermal modeling~\cite{Vinante2019} suggest that temperatures around ∼ $100mK -1K$ could be achievable, with ultralow absorption ($\approx 10^{-8}$ for silica at 1550 nm) and a relatively high gas pressure of $10^{-4}$ mbar~\cite{ahn2018optically}. Furthermore, active refrigeration techniques could be implemented \cite{Rahman2017}. For $T \approx 100 mK -  1K$, use of $\Omega \sim 100MHz - 1 GHz$ would be recommended in order to maintain the thermal signal to noise ratio and $q(\omega_c)$ around ∼ $10^2$. Additionally, since we are looking for an emission at frequency $\omega_c$, detection of such photons will be very prominent against the virtually improbable thermal emission ($\sim e^{-200}$ for $T\sim 100 mK$) at this frequency. 

{\it Atoms:} The atoms required in this experiment must feature electric dipole allowed transitions with frequency of the order of tens of MHz. 
This can be achieved by considering  two levels such as $nS_{1/2}$ and $nP_{1/2}$ of the  Hydrogen atom, which are degenerate in Dirac theory and differ only by the Lamb shift which can be $\sim$ tens of MHz.  Another simple example is the hydrogen atom at a large $n$ and $l$ quantum number. For instance, with $n=10$, one finds an electric dipole allowed transition between $l=9$ and $l=8$ with a frequency of $\sim 100$ MHz. 

{\it Fabrication:} The  foremost concern is protecting the electric dipole transition line while attaching the atom to the rotating silica molecule. This can be ensured using endohedral fullerenes, e.g., using  a C60 fullerene  as a cage to contain the atom \cite{Delaney2004}. The interaction between the cage and the atom is without charge transfer, which leads to negligible change in the encapsulated atom's electronic structure. Next, the fullerene containing the atom is  strongly linked chemically to the nanoparticle either during its synthesis or after to withstand the centrifugal force of rotation. Wet-chemical fullerene chemistry \cite{BOUL1999} as well as ion bombardment techniques \cite{Berning2019} to implant the endohedral fullerene directly into the silica nanoparticle are some of the possible routes for achieving this. Further, as can be seen from  Eq.~\eqref{NIResponse}, noninertial emission effects of the same magnitude  can be obtained through  changing other parameters e.g., using a larger radius of rotation, a larger quality factor, or a smaller volume, for a somewhat slower rotation speed if needed.

{\it Photon detector:} The LC cavity will be monitored by a ultralow noise microwave amplifier, weakly coupled to the LC cavity as usually done in circuit-QED technologies. A crucial benefit of a cryogenic system will be the availability of ultralow noise microwave amplifiers for detecting the photons emitted in the cavity. Here, we take advantage of the cutting edge technology developed in the context of superconducting quantum technologies and circuit-QED, which are optimized for frequencies of the order of $5$ GHz. Josephson parametric amplifiers (JPA) with noise limited only by vacuum fluctuations, i.e., quantum-limited, are at present the best option~\cite{Bergeal2010,Beltran2008}.

Conclusion.  To summarize, we have shown that a rapidly rotating atom inside a far-detuned cavity with experimentally feasible parameters displays a strong signature of noninertial motion through its spectral lines, which are potentially measurable with the current technology. We find that when the cavity is in resonance with the rotational frequency of the atom, which is much larger than its transition frequency, then the dominant contribution to the transition rate comes from the noninertial motion. This setup requires the peak acceleration  $\omega^2R \sim 10^{12} ms^{-2}$ which is  far smaller than the acceleration, on the order of $10^{26}ms^{-2}$ required in other proposals for similar effects.

  % This setup provides a novel avenue where the prediction of noninertial quantum field theory can be cleanly investigated in lab settings, in contrast to all inertial effects. 
 Unlike the case with analog systems, measurement of this kind of radiation will be a direct verification of the concepts of particle creation in quantum fields during noninertial motion. Detection of the modified emission spectrum of the atom due to noninertial motion is not only important for fundamental reasons but it will also boost confidence in many 
quantum field theoretic predictions in geometric backgrounds, e.g., Hawking radiation and gravity-induced particle formation 
\cite{Singleton2011, Steinhauer2014, Steinhauer2016}.

 \section{Appendix}
 \subsection{Detector response in electric field}\label{App:Detector}
% {\em Interaction With Electric Field}\\
We consider  a quantum mechanical (QM) two level atom having proper time $\tau$, interacting with a quantum field $\hat{\phi}$ linearly through a moment $\hat{m}(\tau)$  s.t. $\hat{H}_{\text in}= \hat{m}(\tau) \hat{\phi}[x(\tau)]$. If the initial state of the joint system is $|\phi_i,\psi_i \rangle$ and the final state of interest is $| \phi_f, \psi_f \rangle$ where $|0\rangle, |\phi_f \rangle, |\psi_i \rangle$ and $|\psi_f \rangle$  are the initial state of the field, final state of the field, initial state of the QM system and the final state of the QM system respectively,  the probability amplitude of this transition under first order perturbation theory, in the rest frame of the QM system is
\bea
A^{(1)} =\frac{ i}{\hbar} \langle \phi_f, \psi_f |   \int_{-\infty}^{\infty} d \tau  \hat{m}(\tau) \hat{\phi}[x(\tau)]|\phi_i,\psi_i \rangle.
\eea
If ${\hat H}_0$ is the free Hamiltonian of the QM part, in the interaction picture, we have
$
\hat{m}(\tau) =e^{i\frac{{\hat H}_0 }{\hbar}\tau} \hat{m}(0)  e^{- i\frac{{\hat H}_0 }{\hbar}\tau}.
$
Further the initial   and the final state of the QM system are the ground and excited state with energy $E_0$ and $E_1$ respectively while  the initial state of the field is vacuous and we integrate over all final states of the field we get the probability of transition as 
\begin{equation}
    P_{0\rightarrow 1} =\frac{ |\langle \psi_f |{\hat m}(0) | \psi_i\rangle|^2}{\hbar^2}\iint_{-\infty}^{\infty} d \tau d \tau'e^{-i \Omega (\tau -\tau')}G(x,y),  \label{ProbDef}
  \end{equation}
  where $G(x,y) = \bra{0}\hat{\phi} (x) \hat{\phi}(x') \Ket{0}$ is the Wightman function of the field, while $\Omega =( E_1-E_0)/\hbar$. This, on usage of $\tau_+ =(\tau + \tau')/2$ and $\tau_- = (\tau-\tau')$  leads to a rate of transition defined as $\Gamma_{0\rightarrow 1}  \equiv d P_{0\to 1}/d\tau_+$ 
\bea
\Gamma_{0\rightarrow 1}  = \frac{ |\langle \psi_f |{\hat m}(0) | \psi_i\rangle|^2}{\hbar^2}\int d\tau_-  e^{-i \Omega t_-}  \bra{0}\hat{\phi} (x) \hat{\phi}(x') \Ket{0},\nonumber\\
\eea
 It is seen from this expression that the transition rate $\Gamma$ is proportional to the Fourier transform of the two-point correlation function of the quantum field operator. Due to the fact that quantum correlation of the field is non-zero even in the vacuum state, the above expression gives a non-zero rate of transition for vacuum state as well. This, is a general result true for any kind of field $\hat{\phi}$ interacting with QM system. We will now analyse this in an experimentally more easily realizable atom-electromagnetic field interaction set-up.

An atom in vacuum interacts with the electromagnetic field modes of the vacuum in the ground state.
 The coupling between the electromagnetic modes and the atom is characterized by the interaction Hamiltonian 
$H_{\text{in}}= -\hat d_{\mu}E^{\mu}$, where $\hat d^{\mu}$ and $ E^{\mu}$ are the atomic dipole and the electric field vectors in the rest frame of the atom. If the dipole moment is taken to be along the $z$-axis, then
$H_{\text{in}} = -\hat d E^{z}$ in the inertial frame, where $\hat d^2=\hat d^{\mu}\hat d_{\mu}$ gives the magnitude of the atomic  dipole.  
Thus, for the case where the system is interacting with an electric field with a time dependent dipole moment, the transition rate  becomes,
\begin{align}
 \Gamma_{0\rightarrow 1}  = \frac{1}{\hbar^2}\int d \tau_- |\langle \psi_f |\hat d | \psi_i\rangle|^2 e^{-i \Omega \tau_-}  \bra{0} E^z(x) E^z(x') \Ket{0}.
\end{align}

For the $z-$component of the electric field, the two-point correlation function reads
\begin{widetext}
\begin{align}
  \bra{0} E^z(\bm{x},t) E^z(\bm{x}',t') \Ket{0} = \int \frac{d^3 k \rho_f (k)}{(2 \pi)^6 } e^{- i \omega_k (t-t')}\frac{\omega_k}{2}\left( 1 -\frac{( k^z)^2}{{\bf k }^2}\right)e^{i {\bf k \cdot [x(t)-x(t')]}}, \label{Electric2Pt}
\end{align}
\end{widetext}
where $\rho_f (k)$ is the density of sates in free space.

Similarly in the rest frame of the atom, i.e. $ \Delta x^i =0$, using  \eqref{Electric2Pt}, we can see that there exists a non-zero probability of  inverse transition,  process of de-excitation ($1\rightarrow 0$) given by
  \begin{align}
P_{1\rightarrow 0} = \frac{{d}^2  }{\epsilon_0 \pi ^2 \hbar c^3}\iint_{-\infty}^{\infty}d \tau d \tau' \frac{e^{i \Omega (\tau -\tau')}}{(\tau -\tau')^4},    
  \end{align}
which leads to the process of {\it spontaneous emission} in an excited atom. Therefore, the spontaneous emission rate $\Gamma_0 = d P_{1\to 0}/d\tau_+$ in the rest frame  reads (Each $\tau_- $ (or $t_-$ in the later portions) appearing in the Wightman function is actually understood to be $\tau_- - i \epsilon$  (or  $t_- - i \epsilon$).)
\begin{align}
\Gamma_0 =  \frac{{d}^2 }{\epsilon_0 \pi ^2 \hbar c^3}\int d \tau_-\frac{e^{i \Omega\tau_-}}{\tau_-^4} = \frac{\tilde{d}^2 \Omega^3}{ 3 \pi \epsilon_0  \hbar c^3}. \label{SPEMR}
\end{align}
%Here we have changed the variables from $\tau, \tau'$ to $\tau_+ = (\tau + \tau')/2$ and $\tau_- = \tau - \tau'$.

 Finally before we go to discuss the theoretical construction of the proposed set-up, for completeness, we make note of a rather obvious point. If the atom is moving with velocity $v$ in the lab frame, then the clocks of the atom and the lab are related by  $\tau = \gamma^{-1} t$ and  $(\tau -\tau' )^2 = \gamma^{-2} (t-t')^2 = [ (t-t')^2- (x-x')^2]$, where $\gamma = 1/\sqrt{1-v^2/c^2}$. This  will yield the the spontaneous emission rate in the lab frame $\Gamma_\gamma = \Gamma_0/\gamma$. 
The same can be obtained from first writing the electric field two-point function and then transforming it as a tensor to the lab frame. Similarly, using the co-ordinates of an eternally accelerating Rindler observer \cite{Unruh1976} we obtain the standard thermal response rate due to its acceleration  $\Gamma_T = \Gamma_0 /(1-\exp{-2 \pi c \Omega /a})$.

We will do this exercise for a rotating atom put inside a cavity, in the lab frame. In this case the dipole is pointing in a direction (say $y-$) which is orthogonal to the plane ($x-z$) of the rotation.

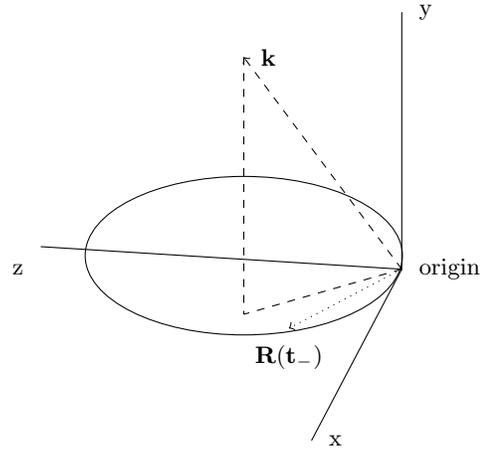
\begin{figure}[h] 
%    \begin{center}
      \begin{tikzpicture}[scale=0.6]
       \draw (-3.5,1.6) ellipse (100pt and 50 pt);
         \draw[-] (0,1.3) -- (0,7);
         \draw[-] (0,1.3) -- (-8,1.8);
          \draw[dotted,->] (0,1.3) -- (-2.5,-0.01);
         \draw[dashed,->] (0,1.3) -- (-3.5, 6);
         \draw[dashed]  (-3.5, 6)--(-3.5,0.3);
          \draw[-] (0,1.3) -- (-2, -2.5);
          \draw[dashed]  (0,1.3)--(-3.5,0.3);
          
          \node[label=right : y](4) at (0,7) {};
          \node[label=right: origin ](5) at (0,1.3) {};
          \node[label=below: ${\bf R(t_-)}$](6) at (-2.5,-0.01) {};
          \node[label=right :{\bf  k}](7) at (-3.5, 6) {}; 
          \node[label=left : z](8) at (-8,1.3) {}; 
          \node[label=right : x](9) at (-2, -2.5) {}; 
        \end{tikzpicture}
 \caption{The vector ${\bf R(t_-)} = {\bf x(t)-x(t')}$ rotates about the $y-$axis while making an angle $\alpha =\omega t_-/2 -\pi/2$ with the $z-$ axis. The projection of a given mode vector ${\bf k}$ has to be evaluated along  ${\bf R(t_-)}$ which itself keeps changing over time.}
    \label{fig:RotatingSpace}
\end{figure}

The co-ordinates for a rotating particle inside the cavity,  are given in the lab frame $(t,x,y,z)$  as
\bea
x(t)= x_0 + R \cos{\omega t},\ z(t)= z_0+R \sin{\omega t},\ y=0,
\eea
where $R$ is the radius of rotation and $\omega$ is the rotation speed .
In a non-inertial frame co-rotating with the atom, the co-ordinates are related as
\begin{align}
x' = x-x_0-R  \cos{\omega t}, \ z'=z-z_0- R \sin{\omega t},\ y'=y.
\end{align}
Thus, the proper time and the co-ordinate (lab) time are related as
\bea
d\tau = \left( 1-\frac{\omega^2 R^2}{c^2}\right)^{\frac{1}{2}}dt \label{t-Transfm}
\eea 
leading to a line element
\begin{widetext}
\bea
ds^2 = -d \tau^2 + \sum_i (dx'_i)^2  -\frac{2 \omega R \sin{\omega t(\tau)} }{ (1-\omega^2 R^2)^{\frac{1}{2}}}d\tau dx' + \frac{2 \omega R \cos{\omega t(\tau)} }{ (1-\omega^2 R^2)^{\frac{1}{2}}}d\tau dz'. \label{linElmt}
\eea
\end{widetext}
Now, in the rest frame of the atom the interaction Hamiltonian is $H_{\text{int}}=-g_{\mu \nu} \hat d'^{\mu}E'^{\nu}$, where the primed quantities are the vectors as seen from the co-moving frame. Using the fact that the dipole 4-vector  is orthogonal to the world line of the static observer (in the rest frame) i.e. $g_{\mu \nu} d'^{\mu} u'^{\nu} =0$, for $u'^{\mu}=(1,0,0,0)$ we get $d'^0 =0$ leading to the fact $d^2 = g_{\mu \nu} d'^{\mu}d'^{\nu} = (d'^y)^2.$  Since in the atom's frame, the dipole is pointing in the $y'$-direction, the interaction Hamiltonian  becomes
\bea
H_{\text{int}}=-g_{\mu \nu} \hat d'^{\mu}E'^{\nu} = - \hat d'^y E'^y.
\eea
Thus, we can transform the interaction term to the lab frame
\bea
H_{\text{int}}=  \frac{-\hat d}{(1-\frac{\omega^2 R^2}{c^2})^{\frac{1}{2}}}\left[ E^y - \frac{ R \omega}{c}\sin{\omega t} c  B^z- \frac{ R \omega}{c} \cos{\omega t} c  B^y  \right]. \nonumber\\
\eea
Therefore, in the first order perturbation theory we end up-getting the two-point correlator of
\begin{widetext}
\bea
\langle 0| \left( E^y (t) -\frac{ R \omega}{c}  \sin{\omega t} c  B^z(t) -\frac{ R \omega}{c} \cos{\omega t} c  B^y  \right)\left( E^y(t') -\frac{ R\omega}{c} \sin{\omega t'} c  B^z(t') -\frac{ R\omega}{c}  \cos{\omega t'} c  B^y (t') \right)|0 \rangle \label{Inertial2-PtTrasnfm}
\eea
\end{widetext}
We also transform the integration measure $d \tau d \tau'$ in \eqref{ProbDef} using \eqref{t-Transfm} and again go to the variables $t_+ = (t +t')/2$ and $t_- = t - t'$. Since the various two point correlators of the electromagnetic fields in inertial vacuum $\langle 0 | E^{i}(t) B^j(t')| 0 \rangle$ depend on $t_-$  the crossed terms of \eqref{Inertial2-PtTrasnfm} vanish under $t_+$ integration. The relevant surviving correlators will be given as
\begin{widetext}
\bea
\langle 0| E^y (t)  E^y(t') |0 \rangle = \int \frac{d^3 k \rho(k)}{(2 \pi)^6 V} e^{- i \omega_k t_-}\frac{\omega_k}{2}\left( 1 -\frac{( k^y)^2}{{\bf k }^2}\right)e^{i {\bf k \cdot (x(t)-x(t'))}},\\
\langle 0| B^y (t) B^y(t') |0 \rangle = \int \frac{d^3 k \rho(k)}{(2 \pi)^6 V} e^{- i \omega_k t_-}\frac{\omega_k}{2 c^2}\left( 1 -\frac{( k^y)^2}{{\bf k }^2}\right)e^{i {\bf k \cdot (x(t)-x(t'))}},\\
\langle 0| B^z (t)  B^z(t') |0 \rangle = \int \frac{d^3 k \rho(k)}{(2 \pi)^6 V} e^{- i \omega_k t_-}\frac{\omega_k}{2c^2}\left( 1 -\frac{( k^z)^2}{{\bf k }^2}\right)e^{i {\bf k \cdot (x(t)-x(t'))}},\\
\eea
\end{widetext}
where $\rho(k)$ is the density of energy states inside the cavity and $V$ its volume.
Lastly using the fact that the vector ${\bf x(t)-x(t')}$ changes both direction and magnitude, $ (x(t)-x(t'))^2 = 4 R^2 \sin^2{\omega t_-/2}$ over time and performing the $t_-$ integration (see Fig. \ref{fig:RotatingSpace})
we get
\begin{widetext}
\bea \label{TransitionRate}
\Gamma_{cav} \sim  \frac{d^2}{\epsilon_0 \hbar} \frac{1}{V}\int_0^{\infty} dk  \rho(k)\omega_k \left[ \underbrace{ \delta(\bar{\Omega} -\omega_k)}_{\text{Inertial contribution}} +\underbrace{\frac{1}{2}\frac{ R^2 \omega^2}{c^2}\{\delta(\bar{\Omega} + \omega -\omega_k )+ \delta(\bar{\Omega}  - \omega - \omega_k )\} }_{\text{ Purely non-inertial contribution}}\right.\nonumber\\
\left.  + \underbrace{\frac{ R^2 \omega_k^2}{c^2} \left( \frac{1}{10} \delta(\bar{\Omega}  -\omega_k ) -\frac{1}{20}\{\delta(\bar{\Omega} + \omega -\omega_k )+ \delta(\bar{\Omega}  - \omega - \omega_k )\} \right)}_{\text{Mixed Contribution}} +\underbrace{{\cal O}\left(\sum_{n>1} \frac{R^{2n} \omega_k^{2n}}{c^{2n}}\delta(\bar{\Omega}  \pm n\omega - \omega_k )\right)}_{\text{Higher resonances}}\right],
\eea
\end{widetext}
with $\bar{\Omega}  =\Omega  \left( 1-\omega^2 R^2/c^2\right)^{1/2} \approx \Omega$. In the expression \ref{TransitionRate}, we have clearly identified various terms in the square bracket inside the integral. The first term is the standard vacuum response, which causes the spontaneous emission in the atom in the inertial frame. The remaining terms become zero in an inertial frame and are dependent on acceleration as well as parameter $\zeta$.  We call the third term in parenthesis as the mixed term as it involves a term having no frequency shift but appearing multiplied with non-inertial parameter $\zeta$. The last term is just due to  the existence of all the higher  resonances the cavity will automatically support. However, these terms come with suppression factors $\zeta$ at higher orders and contribute minutely in the range of parameters discussed in the main draft.

In the limit of $\omega \gg \bar{\Omega} $ we have, 
\bea
\Gamma_{c} \sim \frac{d^2}{\epsilon_0 \hbar} \frac{1}{V} \left[  \bar{\Omega} \left( 1-\frac{\zeta(\bar{\Omega})}{10} \right) \rho( \bar{\Omega})+ \frac{33\zeta(\omega)}{60}\tilde{\omega} \rho(\tilde{\omega} )\right],\label{CavityResponse2}
\eea
where $\tilde{\omega} = \omega + \bar{\Omega}$.
Further, with a Lorentzian density profile with quality factor $Q$ we obtain,
\bea
\rho(k) \xrightarrow{k \sim \omega_c} \frac{Q}{\omega_c}; \hspace{0.1 in}
\rho(k)\xrightarrow{k \ll \omega_c} \frac{1}{\omega_c Q},\hspace{0.1 in} \text{while} \nonumber\\
\rho(k)\xrightarrow{k \gg \omega_c} \frac{\omega_c}{k^2 Q}. \hspace{1.5 in} \label{CavityLims}
\eea
Thus, obtaining
\bea
\Gamma_{c} \sim  \frac{33\zeta(\omega)}{60} \frac{d^2}{\epsilon_0 \hbar} \frac{1}{V} Q.
\eea
for a cavity tuned near $\tilde{\omega}$.

\subsection{Perturbing through a coherent State }\label{App:Coherent}
In a coherent state, such as the laser to be used for trapping/rotating the atom,  the two point function of the vector potential is given as
\begin{align}
\langle C | A^{\mu}(x,t) A^{\nu}(x',t')\Ket{C} =G^{\mu \nu}(x,y)
 +  \langle  A^{\mu}(x,t) \rangle \langle A^{\nu}(x',t')\rangle, \label{CoherentWightman}
\end{align}
where $G^{\mu \nu}(x,y)$ is vacuum two point function~\cite{Tsue1991}. Therefore, the detector response will pick up extra pieces such as
\begin{widetext}
\begin{align}
\int_{-\infty}^{\infty} dt dt' e^{i (\bar{\Omega} \pm \omega)(t-t')} \langle  E^{\mu}(x,t) \rangle \langle B^{\nu}(x',t')\rangle =\int_{-\infty}^{\infty} dt dt' e^{i (\bar{\Omega} \pm \omega)(t-t')}  E_{cl}^{\mu}(t)  B_{cl}^{\nu}(t')
= \tilde{ E}_{cl}^{\mu}(\bar{\Omega}\pm \omega)  \tilde{ B}_{cl}^{\nu}(-\bar{\Omega} \mp \omega),
\end{align}
\end{widetext}
where $ E_{cl}^{\mu},\ \tilde{ E}_{cl}^{\mu}$ etc., are the classical field configuration and their Fourier transforms respectively. Thus unlike the purely non-inertial term, the contribution from the introduction of the coherent state also depends on the distribution profile of the field configuration in the frequency space.  For a monochromatic light very sharply peaked around a frequency $\omega_{laser} \gg \bar{\Omega}\pm \omega$, as is the case with the set-up with 1550 nm light,  the correction term will be much feeble against the vacuum contribution, as the support of the product of Fourier decompositions of the coherent light at $ \bar{\Omega}\pm \omega \ll\omega_{laser}$ will be extremely  minute.

% \section{Acknowledgments}
\begin{acknowledgements}
K.L. acknowledges the financial support from the INSPIRE Faculty fellowship DST INSPIRE (04/2016/000571) from the Government of India.  S.K.G. acknowledges the financial support from research grant ECR/2017/002404 from SERB-DST. H.U. and A.V. acknowledge financial support from the EU H2020 FET project TEQ (Grant No. 766900), the Leverhulme Trust (RPG-2016-046), the COST Action QTSpace (CA15220) and the Foundational Questions Institute (FQXi).  K. L. and S.K.G. wish to thank T. Padmanabhan, Ketan Patel, and Sanjib Dey for useful discussions. 
\end{acknowledgements}

 % \bibliography{Ref}

\end{document}